\documentclass{elsarticle}
\usepackage{amsmath}
\usepackage{graphicx}
\biboptions{sort&compress}
\journal{Physics Letters B}

\begin{document}

\begin{frontmatter}

\title{Fine structure of the isoscalar giant quadrupole resonance in $^{40}$Ca due to Landau damping?}

\author[itl,wits]{I.~Usman}
\author[itl]{Z.~Buthelezi}
\author[wits]{J.~Carter}
\author[witsgeo]{G.R.J.~Cooper}
\author[uct]{R.W.~Fearick}
\author[itl]{S.V.~F\"ortsch}
\author[itl,wits]{H.~Fujita}
\author[osaka]{Y.~Fujita}
\author[ikp]{Y.~Kalmykov}
\author[ikp]{P.~von~Neumann-Cosel\corref{cor1}}
\ead{vnc@ikp.tu-darmstadt.de}
\author[itl]{R.~Neveling}
\author[ikp]{P.~Papakonstantinou}
\author[ikp,ect]{A.~Richter}
\author[ikp]{R.~Roth}
\author[ikp]{A.~Shevchenko}
\author[wits]{E.~Sideras-Haddad}
\author[itl]{F.~D.~Smit}

\cortext[cor1]{Principal corresponding author}

\address[itl]{iThemba LABS, PO Box 722, Somerset West 7129, South Africa}
\address[wits]{School of Physics, University of the
Witwatersrand, Johannesburg 2050, South Africa}
\address[witsgeo]{School of Earth Sciences, University of the
Witwatersrand, Johannesburg 2050, South Africa}
\address[uct]{Department of Physics, University of Cape Town,
Rondebosch 7700, South Africa}
\address[osaka]{Department of Physics, Osaka University, Toyonaka,
Osaka 560-0043, Japan}
\address[ikp]{Institut f\"ur Kernphysik, Technische
Universit\"at Darmstadt, D-64289, Darmstadt, Germany}
\address[ect]{ECT*, Villa Tambosi, I-38123 Villazzano (Trento), Italy}

\begin{abstract}
The fragmentation of the Isoscalar Giant Quadrupole Resonance (ISGQR)
in $^{40}$Ca has been investigated in high energy-resolution
experiments using proton inelastic scattering at $E_{\rm{p}} =
200$~MeV. Fine structure is observed in the region of the ISGQR and its characteristic energy scales are extracted from the experimental data
by means of a wavelet analysis. The experimental scales are well
described by Random Phase Approximation (RPA) and second-RPA
calculations with an effective interaction derived from a realistic
nucleon-nucleon interaction by the Unitary Correlation Operator Method
(UCOM). In these results characteristic scales are already present
at the mean-field level pointing to their origination in Landau damping, in contrast to the findings in heavier nuclei and also to
SRPA calculations for $^{40}$Ca based on phenomenological effective interactions, where fine structure is explained by the coupling to two-particle two-hole (2p-2h) states.
\end{abstract}


\end{frontmatter}

\section{Introduction}\label{intro}

High energy-resolution experiments on nuclei excited in the region of
giant resonances present a unique way to extract information about the dominant processes of the decay. This information is imprinted into the fragmentation of giant resonances visible as fine structure in
excitation energy spectra. Recent work has established fine structure
as a global phenomenon in medium-mass to heavy nuclei for the case of
the Isoscalar Giant Quadrupole Resonance (ISGQR) \cite{she09}. Fine
structure has also been demonstrated for a variety of other modes like
the spin-isospin flip Gamow-Teller (GT) mode \cite{kal06}, the
Isovector Giant Dipole Resonance (IVGDR) \cite{die94,str00,tam07} and
the magnetic quadrupole resonance \cite{vnc99}.

A variety of theoretical approaches has been put forward to understand
the fine structure including doorway-state analysis \cite{win83}, a
local scaling dimension model \cite{aib99}, an entropy index method
\cite{lac99} and  wavelet analysis \cite{she04,hei10}. A comparison for representative cases indicates that a wavelet analysis is a
particularly promising tool \cite{she08}, since it provides
simultaneously a quantitative measure of the fine structure and
information on the localization in the excitation energy spectrum.
Characteristic scales can be extracted from the power spectra of
wavelet transforms, which allow a direct comparison between experiment
and theory. A systematic study of the ISGQR \cite{she09} shows that the observed scales in medium-mass to heavy nuclei originate from a collective damping mechanism induced by the coupling of elementary
one-particle one-hole (1p-1h) states to low-lying surface vibrations
\cite{ber83}.

Recently, the high energy-resolution measurements of the ISGQR have
been extended \cite{usm09} to the low-mass region $12 \le A \le 40$.
These experiments were motivated by the question whether
collective damping remains the most important decay mechanism in light
nuclei or whether other contributions like direct decay or Landau
damping become more important \cite{har01}. The present work focuses on the case of $^{40}$Ca where considerable fragmentation of E2 strength has been observed in electron \cite{die94}, proton \cite{lis89,sch01} and
alpha scattering \cite{bor81} studies, albeit measured with varying
resolution. The choice of a doubly magic nucleus allows a comparison to calculations within the framework of RPA and SRPA and thus to
distinguish between the role of 1p-1h and 2p-2h excitations. Since in
the calculations a modern realistic interaction derived by the Unitary
Correlation Operator Method (UCOM) described in
Refs.~\cite{fel98,rot05,rot10} has been used, the experimental data
serve at the same time as a test bench for UCOM. This interaction is
one of a family of realistic interactions and has
been used with quite some success in the description of basic nuclear quantities such as masses \cite{rot06} and the gross features of giant
resonances, in particular the giant dipole and giant quadrupole ones on the SRPA level \cite{pap09,pap10}.

\section{Experiment}
\label{sec:experiment}

\begin{figure}[tbh]
\centerline{
\includegraphics[width=8.5cm,angle=0]{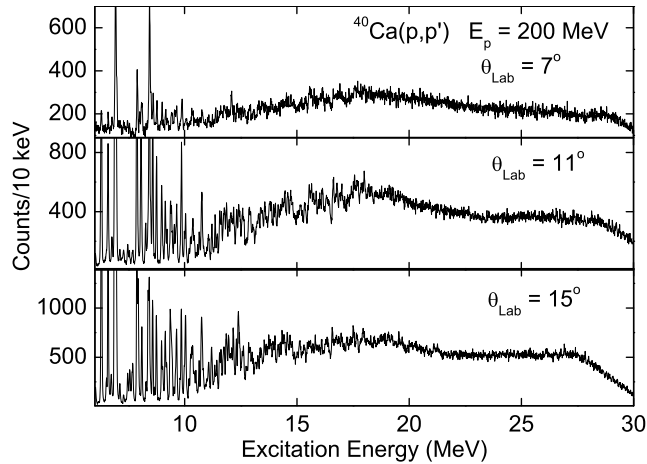}
}
\caption{Excitation energy spectra for $^{40}$Ca at scattering angles
of $\theta_{\rm{Lab}}$ = 7$^{\circ}$, 11$^{\circ}$ and 15$^{\circ}$.
Note that the ISGQR is most strongly excited at $\theta_{\rm{Lab}}$ = 11$^{\circ}$.
The drop of counts in the spectra above 27~MeV is due to a gradual loss of angular acceptance towards the end of the spectrometer focal plane.}
\label{fig:targetspectra}
\end{figure}

The experiments were carried out with a 200~MeV proton beam produced by the Separated Sector Cyclotron (SSC) of the iThemba Laboratory for
Accelerator Based Sciences (iThemba LABS), South Africa. The protons
scattered inelastically on a $^{40}$Ca target with an areal density of
3.0 mg/cm$^{2}$ were momentum analyzed with a K600 magnetic
spectrometer after scattering. Dispersion matching techniques were
used in order to exploit the high energy-resolution capability of the
spectrometer. Energy resolutions $\Delta E = 35 - 40$ keV Full Width at Half Maximum (FWHM) were achieved. The scattering angles were selected
to be below, at and above the maximum of the cross section for $\Delta
L = 2$ transitions into the ISGQR. Therefore, measurements were
performed at three different scattering angles, $\theta_{\rm{Lab}} =
7^\circ$, 11$^\circ$ and 15$^\circ$, as illustrated in
Fig.~\ref{fig:targetspectra}, for excitation energies between 6 and 30
MeV. Details of the data analysis procedures are described elsewhere
\cite{she09}.

The excitation energy spectrum in Fig.~\ref{fig:targetspectra} taken at $\theta_{\rm{Lab}}$ = 11$^{\circ}$, which corresponds to the maximum
cross section for the ISGQR, reveals a broad resonance at a mean energy $E_{\rm{x}} \approx 18$ MeV with strength distributed between approximately 12 and 22~MeV. Intermediate structure is visible through peaks around 12, 14, 16, 17 and 18 MeV consistent with previous experimental work \cite{die94,lis89,sch01,bor81}. Pronounced fine structure is observed up to about 20 MeV. At the larger scattering angle of $\theta_{\rm{Lab}} = 15^{\circ}$, the overall structure is reduced in size but distinctive features of the fine structure persist.
At the smaller scattering angle $\theta_{\rm{Lab}}$ = 7$^{\circ}$, the
fine structure changes somewhat indicating the presence of other multipoles.

The dominant quadrupole nature of the bump on top of a smooth quasifree background \cite{bak97} has been demonstrated by multipole decompositions of angular distributions measured in alpha \cite{you01} and proton \cite{lis89} scattering. Properties of the ISGQR have also been investigated with electromagnetic probes \cite{die94,gor67} and capture experiments \cite{wel80}. The results are consistent in that the energy weighted sum rule for isoscalar E2 strength is almost exhausted in the excitation range of interest.  Strong transitions at energies $E_{\rm x} \leq 12$~MeV consist of a mixture of different multipoles \cite{cam04}, while the peaks in the excitation region around 14~MeV were shown to have quadrupole character \cite{yam87}.

The multipole analysis of proton scattering data in Ref.~\cite{lis89} indicates that for the kinematics of the present experiment at the optimum angle for $\Delta L = 2$ transfer, contributions from the overlapping isoscalar giant monopole resonance (ISGMR) and low-energy $\Delta L = 3,4$ strength are at the level of a few percent at best. The IVGDR is strongly populated at forward scattering angles by relativistic Coulomb excitation. However, the angular distribution falls off steeply such that the contribution is already small (but still sizable) at $\theta_{\rm{Lab}} =7^\circ$ and negligible at $11^\circ$.
Furthermore, an analysis of the spectra along the lines described in Ref.~\cite{kal06} with the additional constraint of a constant level density independent of the kinematics indicates only limited contributions to the cross sections in the ISGQR region from quasifree processes and/or other multipoles.
Thus, it seems justifed to assume that the fine structure of the cross sections at $\theta_{\rm{Lab}} = 11^\circ$ arises solely from the ISGQR.

Another way to verify this conjecture is to use
a cross correlation analysis as described
e.g.\ in Ref.~\cite{car01}. For that purpose, any smooth background in the spectra was removed by a discrete wavelet analysis \cite{she08,kal06}. In a next step gross structures were taken out by folding with a Gaussian of appropriate width. The ratio between these two spectra - the so-called
stationary spectrum - then contains local fluctuations around the mean only \cite{han90}. The normalized cross correlation functions $C_{{\Theta_1},{\Theta_2}}$ between stationary spectra including the $7^\circ$ angle are quite similar: $C_{7^\circ,11^\circ} = 0.26(8)$ and $C_{7^\circ,15^\circ}= 0.21(7)$.
In contrast, a much larger value $C_{11^\circ,15^\circ} = 0.50(8)$ is found for the $11^\circ,15^\circ$ angle pair indicating a high degree of correlation, since the value is above an expression derived for the angular cross-correlation function of cross-section fluctuations \cite{bri64}. We conclude that the fine structure observed in the $11^\circ$ spectrum is indeed due to the ISGQR and contributions from other modes must be small.

\section{Extraction of scales}
\label{sec:scales}
\begin{figure}[tbh]
\centerline{
\includegraphics[width=8.5cm,angle=-90]{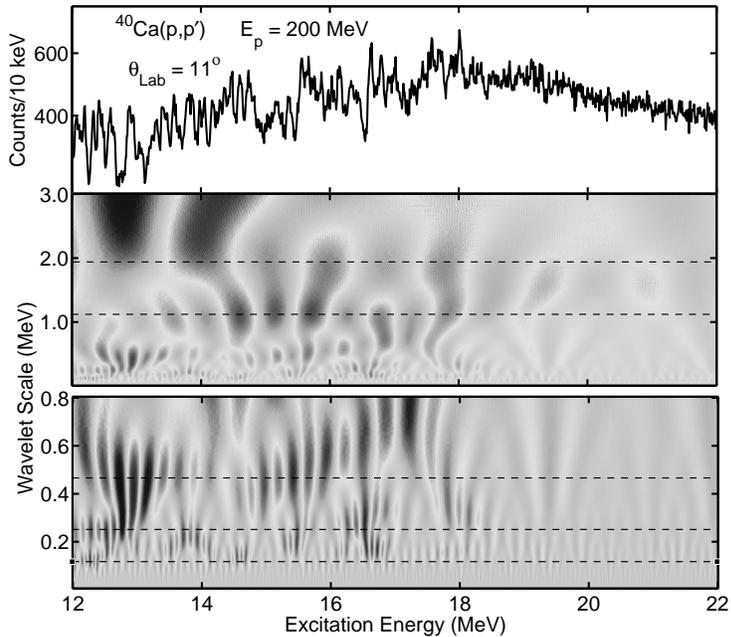}
}
\caption{Excitation energy spectrum and wavelet transform for $^{40}$Ca$(p,p')$ at
$\theta_{\rm{Lab}}$ = 11$^\circ$ using the complex Morlet mother wavelet for
scales up to 3.0 MeV. The bottom part shows the presence of characteristic
scales in the fine structure below 800 keV. Characteristic scales identified by maxima in the power spectrum are indicated by dashed lines.} \label{fig:cawavelet}
\end{figure}

Wavelet analysis techniques \cite{mal98} have been utilized for the
extraction of characteristic energy scales from the region of the ISGQR
excitation \cite{she09}. Continuous Wavelet Transforms (CWT) using both
the real and complex Morlet mother wavelets have been successfully
applied previously \cite{she08}. Application of the CWT to the present
data in the energy region of the ISGQR is shown in
Fig.~\ref{fig:cawavelet}. By plotting the real part of the complex
coefficients in a two-dimensional distribution of energy scales
vs.\ excitation energy, the positions of the structures within
the original energy spectrum can be identified. Maxima of the wavelet
coefficients at certain scale values over the energy region of ISGQR (or
parts of it) indicate characteristic scales. This is illustrated in the
middle part of Fig.~\ref{fig:cawavelet}, where the plotted wavelet
energy scale was restricted to 3.0 MeV in order to reveal scales
related to intermediate structure. The bottom part of
Fig.~\ref{fig:cawavelet} shows the same results but with an expanded
wavelet scale to emphasize the presence of characteristic scales in the
fine structure as well.

\begin{figure}[tbh]
\centerline{
\includegraphics[width=8.5cm,angle=-90]{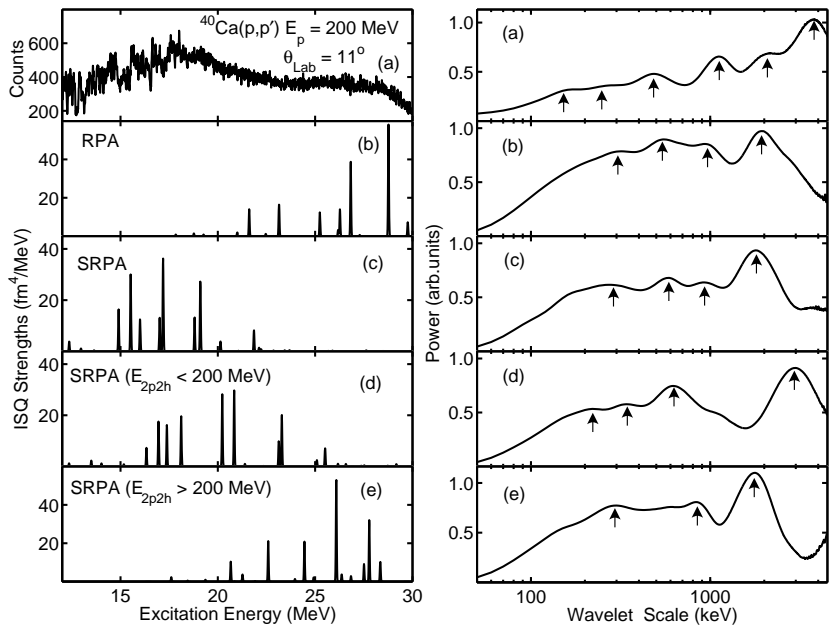}
}
\caption{L.h.s.:Experimental spectrum of the $^{40}$Ca(p,p$^{\prime}$) reaction at the maximum cross section of the ISGQR (a) and theoretical isoscalar quadrupole strength distributions in $^{40}$Ca from RPA (b), full SRPA (c), SRPA including 2p-2h transitions below
200 MeV only (d) and SRPA including 2p-2h transitions above
200 MeV only (e), based on the UCOM interaction. R.h.s.: Corresponding power spectra from the CWT. Characteristic scales are indicated by arrows.}
\label{fig:catheory}
\end{figure}

In order to obtain a quantitative measure for the characteristic scales, the absolute values of the complex coefficients are projected onto the wavelet scale axis.
The resulting power spectrum obtained is shown
in r.h.s.\ of Fig.~\ref{fig:catheory}(a) and maxima representing characteristic scales are indicated by arrows.
Three peaks at scale values 150, 240 and 460~keV arising from the fine structure can be identified as well as three scales at 1.05, 2.0 and 3.9~MeV corresponding to intermediate structure.

\section{Model calculations and discussion}\label{sec:theory}

To the best of our knowledge, all existing RPA calculations of the
ISGQR in $^{40}$Ca (see Refs.~\cite{ter07,nak09,cao09} for some recent
examples) except the one discussed below and shown in Fig.~\ref{fig:catheory}(b) on the l.h.s.\ are characterized by a dominant peak, which cannot yield a
scale in the wavelet power spectrum (cf.\ Fig.~13 in
Ref.~\cite{she09}).
We have verified this result using Skyrme and Gogny interactions.
Thus, the experimentally observed fragmentation is
usually attributed to mixing of the 1p-1h states with more complex configurations,
including collective phonons \cite{bro81,dro90,spe91,kam03,lac04}.

As such, here we perform large-scale RPA and SRPA calculations with a realistic
interaction, $V_{\mathrm{UCOM}}$. The main differences of the present to other SRPA approaches are:
(i) The two-body, finite-range interaction is the sole input,
describing both the ground state and the residual couplings. In
particular, we use $V_{\mathrm{UCOM}}$, a potential derived
from the Argonne V18 interaction by renormalizing it within the Unitary
Correlation Operator Method \cite{rot05,rot10}. (ii) No truncation is
imposed on the model space describing the excited states, other than
that due to the exhaustion of the single-particle space. This means
that the Hartree-Fock (HF) equations are solved within a selected
single-particle space and all 1p-1h and 2p-2h configurations available
within that space are taken into account when building the SRPA matrix.
The single-particle basis used here consists of harmonic-oscillator
wave functions with orbital angular momentum $\ell\leq 6$ and node
quantum number $n\leq 6$, providing reasonable convergence. The number
of 1p-1h and 2p-2h configurations is about $10^6$. Further details on
the SRPA approach can be found in Refs.~\cite{pap09,pap10}.

The isoscalar quadrupole strength distributions and power spectra resulting from
both RPA and SRPA are shown in Figs.~\ref{fig:catheory}(b) and (c),
respectively.
While the RPA results using $V_{\mathrm{UCOM}}$ sytematically overestimate ISGQR energies \cite{paa06}, a striking result is that the ISGQR strength in $^{40}$Ca already appears
fragmented at the RPA level. The same interaction produces one dominant ISGQR peak for heavier nuclei like $^{90}$Zr and $^{208}$Pb \cite{paa06}. We return to this point later.
In SRPA the resonance is shifted strongly to lower energies compared to
RPA. Indeed, since $V_{\mathrm{UCOM}}$ is a bare interaction with
respect to long-range correlations, the SRPA is employed not only to
supplement RPA with a damping mechanism, but also to effectively dress
the 1p-1h configurations, affecting their energy (for a relevant
discussion see Refs.~\cite{pap09,pap10}). Also observed is that the
resonance strength and the profile of the strength distribution are
altered, but, as is usually the case with this type of calculation \cite{pap10,gam10}, not dramatically.
The SRPA with the present interaction reproduces the
global properties of the ISGQR well
(the same holds for the IVGDR),
considering that the input has not
been fitted to this end \cite{pap10}. In this work it is used for the
first time to analyze the fine structure of the ISGQR.

The power spectra analysis reveals that the
intermediate and large scales are comparable in RPA and SRPA, but in
SRPA small scales are somewhat more important in the power spectrum.
Characteristic scales indicated by arrows in Fig.~\ref{fig:catheory} show a good correspondence between the theoretical results (for
both RPA and SRPA) and the experimental data except for the smallest scale where the relation to the calculations is unclear,
and the largest one, which is suppressed in the calculations.

The question arises as to the origin of the various scales.
The RPA model accounts only for Landau damping.
Therefore, our RPA results, Fig.~\ref{fig:catheory}(b),
suggest that this mechanism plays an important role in the case of
$^{40}$Ca (but not in heavier nuclei),
contrary to predictions using
phenomenological effective interactions and density functionals, as already mentioned.
Regarding the $|ph^{-1}\rangle$ structure of the dominant peaks, we find that the leading components as well as their spacings are quite similar in RPA and SRPA. For example, the two strongest peaks at the higher-energy part of the resonance (RPA: $26.81,28.75$~MeV; SRPA: $17.17,19.09$~MeV)
in both cases contain large $|0g,0d^{-1}\rangle$ components,
while the lowest prominent peak (RPA: 21.59~MeV; SRPA: 14.88~MeV) in both cases is rather dominated by $|1d,(0d \, \mathrm{or} \, 1s)^{-1}\rangle$ configurations.

Apparently, Landau damping accounts for most of the scales.
This result must be a consequence of the properties of the $V_{\mathrm{UCOM}}$, which
differs in many respects from phenomenological effective interactions.
Firstly, it retains all the complexities of the realistic NN
interaction and has not been fitted using HF and RPA calculations
(therefore it yields, e.g., underbound nuclei in HF and too high GQR
energies in RPA calculations).
We found strong Landau fragmentation
also with a different renormalized realistic interaction, namely the
CD-Bonn potential transformed via the Similarity Renormalization Group
(SRG), suggesting that we are not observing an artifact of the UCOM
procedure.
Semi-realistic effective interactions supplemented with a phenomenological three-body contact term~\cite{gue10},
leading to an improved description of nuclear radii~\cite{gue10} and of the three major giant resonances on the RPA level~\cite{guexx},
still produce considerable fragmentation, by comparison.
Regarding the tensor terms of $V_{\mathrm{UCOM}}$ which do
not affect the HF ground state, we find only a weak influence on the
RPA strength function, although in SRPA they account for roughly half
of the energy shift of the ISGQR between RPA and SRPA results.
Secondly, the present interaction does not parameterize long-range
correlations and corresponds to a very low nucleon effective mass
$m^{\ast}/m$ in HF, as is evident from the very broad unperturbed
spectra~\cite{rot10,paa06}.
This alone, however, seems not to be a critical property. We find, for example, that
for traditional Skyrme functionals and using continuum RPA,
a low $m^{\ast}/m$ value does not lead to fragmentation.
Neither does the coupling with the particle continuum produce sufficient spreading (only up to about 1~MeV FWHM).

Next, we examine the role of 2p-2h configurations in the vicinity of
the resonance and of high-lying ones (the present model space contains
2p-2h configurations with energies up to about 600~MeV). In
Figs.~\ref{fig:catheory}(d) and (e), SRPA results are shown using only
2p-2h configurations with energies below or above 200~MeV, respectively, the latter
being roughly the median 2p-2h energy within the present model space.
The configurations in the vicinity of the resonance alone tend to
produce spreading, but the corresponding scales are somewhat too large and the energy of the resonance is also too high, as expected
\cite{pap10}. Subsequent inclusion of the highest states compresses
and shifts the spectrum to lower energies. Conversely, high-lying 2p-2h
configurations alone tend to produce smaller scales.

The smallest SRPA scale must reflect, to some extent, the structure of
the density of states in the vicinity of the resonance and in
particular the distribution of 2p-2h configurations acting as
doorway states. Therefore, its precise value and prominence could be sensitive to the
diagonal approximation (use of unperturbed 2p-2h configurations) employed in the
present SRPA calculation. That approximation is expected to have no
other significant influence on our results though~\cite{pap09,pap10}.

Within the Quasiparticle Phonon Model (QPM) and the Extended Time
Dependent Hartree-Fock (ETDHF) model, it has been possible to attribute scales observed in the ISGQR to the coupling to low-lying surface
vibrations (collective mechanism) \cite{she09,she04}. In
the spirit of Refs.~\cite{she09,she04}, we have attempted to solve the
SRPA problem while excluding or selecting the strongest matrix
elements coupling the 1p-1h and the 2p-2h spaces.  The probability
distribution of the matrix elements here is similar to QPM results
(strongly cusp-shaped around zero). No conclusion could be reached as
to the effect of the few strongest or the many weak couplings. Small
scales were present in all cases. We note that the present formalism
contains only incoherent couplings to the 2p-2h space,
although some collective-phonon effects may be present on a
Tamm-Dancoff approximation level, to the extent that the diagonal
approximation has been found to be rather good. In any case, it is not
possible to isolate collective effects unambiguously, if at all.

No clear signature is found in the calculatons for the smallest experimental scale. Fine structure in the experimental spectra at these scales is most likely due to overlapping Compound nuclear states, i.e.\ Ericson fluctuations \cite{eri63} damped by the experimental resolution \cite{car01}.
Finally, regarding the largest experimentally extracted scale at about 4~MeV, we can only speculate that it is due to the smooth part of the resonance (its width), not reproduced by the calculations.
The smooth part certainly contains contributions due to a coupling with the particle continuum, broadening all peaks. Escape widths in light nuclei up to $^{40}$Ca have been shown to be significant \cite{har01}. Also higher-order configurations and correlations may play a role.
We note that in some existing SRPA and extended shell-model calculations empirical single-particle states are used, i.e., the input is already correlated which may be the reason why they appear more efficient in producing more or less a smooth spreading~\cite{dro90,spe91,sch10}.
This discussion, however, is beyond the scope of the present work.

\section{Conclusions}\label{sec:conclusion}

Fine structure of the ISGQR in $^{40}$Ca has been observed in high
energy-resolution proton scattering experiments and analyzed in the
framework of a wavelet approach. Comparisons with RPA and SRPA
calculations using the UCOM interaction reveal that the scales
characterizing the fine structure are largely present already on the
mean-field level. This finding is in contrast to previous calculations
of the ISGQR strength distribution in $^{40}$Ca, which attribute the
fragmentation to the coupling of 1p-1h states to low-lying phonons \cite{bro81,dro90,spe91,kam03,lac04}.
At present we cannot exclude that the agreement of scales deduced from RPA and SRPA is - at least partly - fortuitous. However,
tests using different interactions indicate that the above result is possibly
a consequence of the properties of the renormalized interaction
used, which retains all the complexities of the realistic NN
interaction as compared to phenomenological mean-field interactions.
The dressing of 1p-1h states in a SRPA calculation is necessary to
achieve a proper reproduction of the mean ISGQR energy and to improve
details of the strength distribution.
It seems unlikely that a full treatment of the particle continuum can lead to fragmentation.
At the same time, we cannot rule out
that coupling to collective phonons additionally plays a role in generating the experimentally observed scales.
The present results underline the
relevance of studies of the fine structure of giant resonances as a
tool for an improved understanding of collective modes and their decay.

\section*{Acknowledgements}

We are indebted to L.~Conradie and the accelerator crew at iThemba LABS for providing excellent beams. Useful discussions with G.~Col\`{o},
D.~Lacroix and P.G.~Rein\-hard are acknowledged. This work has been supported by the DFG under contracts SFB 634 and NE 679/2-2, and by the South African NRF.

\section*{Note added in proof}

Recent work by Aiba et al.~\cite{aib11}, published after submission of our manuscript, circumstantiates our conclusions based on an extraction of the local scaling dimension of fine structure fluctuations in shell-model calculations of the $^{40}$Ca GQR strength function.

\end{document}